\documentclass[letterpaper,twocolumn,10pt]{article}

\usepackage{usenix,epsfig,endnotes}

\usepackage{pifont}
\usepackage{pbox}
\usepackage{hyperref}
\usepackage{listings}
\usepackage[table]{xcolor}

\usepackage[T1]{fontenc}
\usepackage[scaled=0.8]{beramono}

\usepackage[defaultsans,scale=0.9]{cantarell}

\usepackage[sortcites]{biblatex}
\definecolor{mygray}{rgb}{0.95,0.95,0.95}
\definecolor{darkblue}{rgb}{0,0,0.7}
\hypersetup{
	colorlinks=true,
	urlcolor=darkblue,
	linkcolor=darkblue,
	citecolor=darkblue
}

\begin{document}

\title{
	\Large \bf
	Spoiled Onions:\\Exposing Malicious Tor Exit Relays
}

\author{
	{\rm Philipp Winter} \\
	Karlstad University
	\and
	{\rm Stefan Lindskog} \\
	Karlstad University
}

\maketitle

\rowcolors{1}{white}{mygray}

\begin{abstract}
Several hundred Tor exit relays together push more than 1 GiB/s of network traffic.  However, it is
easy for exit relays to snoop and tamper with anonymised network traffic and as all relays are run
by independent volunteers, not all of them are innocuous.
In this paper, we seek to expose malicious exit relays and document their actions.  First, we
monitored the Tor network after developing a fast and modular exit relay scanner.  We implemented
several scanning modules for detecting common attacks and used them to probe all exit relays over a
period of four months.  We discovered numerous malicious exit relays engaging in different attacks.
To reduce the attack surface users are exposed to, we further discuss the design and implementation
of a browser extension patch which fetches and compares suspicious X.509 certificates over
independent Tor circuits.
Our work makes it possible to continuously monitor Tor exit relays.  We are able to detect and
thwart many man-in-the-middle attacks which makes the network safer for its users.  All our code is
available under a free license.
\end{abstract}

\section{Introduction}

\begin{figure}[t]
	\centering
	\includegraphics[width=0.42\textwidth]{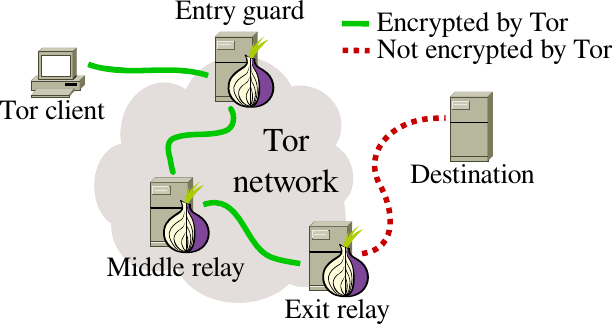}
	\caption{The structure of a three-hop Tor circuit.  Exit relays constitute the bridge between
	encrypted circuits and the open Internet.  As a result, exit relay operators can see---and
	tamper with---the anonymised traffic of users.}
	\label{fig:exits}
\end{figure}

As of January 2014, nearly 1,000 exit relays~\cite{metrics} distributed all around the globe serve
as part of the Tor anonymity network~\cite{Dingledine2004}.  As illustrated in
Figure~\ref{fig:exits}, the purpose of these relays is to establish a bridge between the Tor network
and the ``open'' Internet.  A user's Tor circuits, which are encrypted tunnels, terminate at exit
relays and from there, the user's traffic proceeds to travel over the open Internet to its final
destination.  Since exit relays can see traffic as it is sent by a Tor user, their role is
particularly sensitive compared to entry guards and middle relays; especially because traffic
frequently lacks end-to-end encryption.

By design, exit relays act as a ``man-in-the-middle'' (MitM) in between a user and her
destination.  This renders it possible for exit relay operators to run various MitM attacks such as
traffic sniffing, DNS poisoning, and SSL-based attacks such as HTTPS MitM and
sslstrip~\cite{sslstrip}.  An additional benefit for attackers is that exit relays can be set up
quickly and anonymously, making it very difficult to trace attacks back to their origin.  While it
is possible for relay operators to specify contact information such as an email
address\footnote{Contact information can be useful to get in touch with relay operators, e.g., if
they misconfigured their relay.}, this is optional.  As of January 2014, only 56\% out of all 4,962
relays publish contact information.  Even fewer relays have \emph{valid} contact information.

To thwart a number of popular attacks, TorBrowser~\cite{torbrowser}---the Tor Project's modified
version of Firefox---ships with extensions such as HTTPS-Everywhere~\cite{httpseverywhere} and
NoScript~\cite{noscript}.  While HTTPS-Everywhere provides rules to rewrite HTTP traffic to HTTPS
traffic, NoScript attempts to prevent many script-based attacks.  However, there is little users can
do if web sites implement poor security such as the lack of site-wide TLS, session cookies being
sent in the clear, or using weak cipher suites in their web server configuration.  Often, such bad
practices enable attackers to spy on users' traffic or, even worse, hijack accounts.  Besides,
TorBrowser cannot protect against attacks targeting protocols such as SSH.

All these attacks are not just of theoretical nature.  In 2007, a security researcher published 100
POP3 government credentials he captured by sniffing traffic on a set of exit relays under his
control~\cite{egerstad}; supposedly to show the need for end-to-end encryption when using Tor.  In
Section~\ref{sec:related}, we will discuss additional attacks which were found in the wild.



\subsection{What Happens to Bad Exits?}
The Tor Project has a way to prevent clients from selecting bad exit relays as the last hop in their
three-hop circuits.  After a suspected relay is communicated to the project, the reported attack is
first reproduced.  If the attack can be verified, a subset of two (out of all nine) directory
authority operators manually blacklist the relay using Tor's \texttt{AuthDirBadExit} configuration
option.  Every hour, the directory authorities vote on the \emph{network consensus} which is a
signed list of all relays, the network is comprised of.  Among other information, the consensus
includes the \emph{BadExit flag}.  As long as the majority of the authorities responsible for the
BadExit flag, i.e., two out of two, agree on the flag being set for a particular relay, the next
network consensus will label the respective relay as BadExit.  After the consensus was then signed
by a sufficient number of directory authorities, it propagates through the network and is eventually
used by all Tor clients after a maximum of three hours.  From then on, clients will no longer select
relays labelled as BadExit as the last hop in their circuits.  Note that this does not mean that
BadExit relays become effectively useless.  They keep getting selected by clients as their entry
guards and middle relays.  All the malicious relays we discovered were assigned the BadExit flag.

Note that the BadExit flag is not only given to relays which are proven to be malicious.  It is also
assigned to relays which are misconfigured or are otherwise unable to fulfil their duty of providing
unfiltered Internet access.  A frequent cause of misconfiguration is the use of third-party DNS
resolvers which block certain web site categories.

Apart from the BadExit flag, directory authorities can blacklist relays by disabling its
\emph{Valid} flag which prevents clients from selecting the relay for \emph{any} hop in its circuit.
This option can be useful to disable relays running a broken version of Tor or are suspected to
engage in end-to-end correlation attacks.



\subsection{Contributions}
The three main contributions of this paper are as follows.
\begin{itemize}
	\item We discuss the design and implementation of \textsf{exitmap}; a flexible and fast exit
		relay scanner which is able to detect several popular MitM attacks.
	\item Using \textsf{exitmap}, we monitored the Tor network over a period of four months.  We
		analyse the attacks we discovered in the wild during that time period.
	\item We propose the design and prototype of a browser extension patch which fetches and
		compares X.509 certificates over diverging Tor circuits.  That allows our patch to detect
		MitM attacks against HTTPS.
\end{itemize}

The remainder of this paper is structured as follows.  Section~\ref{sec:related} begins by giving an
overview of related work.  It is followed by Section~\ref{sec:scanning} which discusses the design
and implementation of \textsf{exitmap}.  Section~\ref{sec:results} then presents the attacks we
discovered in the wild.  Next, Section~\ref{sec:thwarting} proposes the design and implementation of
a browser extension patch which can protect against HTTPS MitM attacks.  Finally,
Section~\ref{sec:conclusions} concludes this paper.

\section{Related Work}
\label{sec:related}
While MitM attacks have generally received considerable attention in the
literature~\cite{Holz2012,Wendlandt2008}, their occurrence in the Tor network remains largely
unexplored.  This is unfortunate as the Tor network enables the study of real-world MitM attacks
which are rare and poorly documented outside the Tor network.

In 2006, Perry began developing the framework ``Snakes on a Tor'' (\textsf{SoaT})~\cite{soat}.
\textsf{SoaT} is a Tor network scanner whose purpose---similar to our work---is to detect
misbehaving exit relays.  Decoy content is first fetched over Tor, then over a direct Internet
connection, and finally compared.  Over time, \textsf{SoaT} was extended with support for HTTP,
HTTPS, SSH and several other protocols.  However, \textsf{SoaT} is no longer maintained and makes
use of deprecated libraries.  Compared to \textsf{SoaT}, our design is more flexible and
significantly faster.

Similar to \textsf{SoaT}, Marlinspike implemented \textsf{tortunnel}~\cite{tortunnel}.  The tool
exposes a local SOCKS interface which accepts connections from arbitrary applications.  Incoming
data is then sent over exit relays using one-hop circuits.  By default, \textsf{exitmap} does not
use one-hop circuits as that could be detected by attackers which could then act innocuously.

A first attempt to detect malicious exit relays was made in 2008 by McCoy \emph{et
al.}~\cite{McCoy2008}.  The authors established decoy connections to servers under their control.
They further controlled the authoritative DNS server responsible for the decoy hosts' domain names.
As long as an attacker on an exit relay sniffed network traffic with reverse DNS lookups being
enabled, the authors were able to map reverse lookups to exit relays by monitoring the authoritative
DNS server's traffic.  Using that side channel, McCoy \emph{et al.} were able to find one exit relay
sniffing POP3 traffic at port 110.  However, attackers could avoid that side channel by disabling
reverse lookups.  The popular tool \textsf{tcpdump} implements the command line switch \texttt{-n}
for that exact purpose.

In 2011, Chakravarty \emph{et al.}~\cite{Chakravarty2011} attempted to detect exit relays sniffing
Tor users' traffic by systematically transmitting decoy credentials over all active exit relays.
Over a period of ten months, the authors uncovered ten relays engaging in traffic snooping.
Chakravarty \emph{et al}. could verify that the operators were sniffing exit traffic because they
were later found to have logged in using the snooped credentials.  While the work of Chakravarty
\emph{et al.} represents an important first step towards monitoring the Tor network, their technique
only focused on SMTP and IMAP.  At the time of writing, only 20 out of all $\sim$1,000 exit relays
allow exiting to port 25.  HTTP appears to be significantly more popular~\cite{McCoy2008,Huber2010}.
Also, similar to McCoy \emph{et al.}, the authors only focused on traffic snooping attacks which are
passive.  Active attacks remain entirely unexplored until today.

The Tor Project used to maintain a web page documenting misbehaving relays which were assigned the
BadExit flag~\cite{badexits}.  As of January 2014, this page lists 35 exit relays which were
discovered in between April 2010 and July 2013.  Note that not all of these relays engaged in
attacks; almost half of them ran misconfigured anti virus scanners or used broken exit
policies\footnote{An exit relay's \emph{exit policy} determines to which addresses and ports the
relay forwards traffic to.  Often, relay operators choose to not forward traffic to well-known file
sharing ports in order to avoid copyright infringement.}.

Since Chakravarty \emph{et al.}, no systematic study to spot malicious exits was conducted.  Only
some isolated anecdotal evidence emerged~\cite{furry}.  Our work is the first to give a
comprehensive overview of \emph{active attacks}.  We further publish our code under a free
license\footnote{See: \url{http://www.cs.kau.se/philwint/spoiled_onions}.}.  By doing so, we enable and
encourage continuous and \emph{crowd-sourced} measurements rather than one-time scans.

\section{Probing Exit Relays}
\label{sec:scanning}
We now discuss the design and implementation of \textsf{exitmap} which is a lightweight Python-based
\emph{exit relay scanner}.  Its purpose is to create custom circuits to exit relays which are then
probed by modules which establish decoy connections to various destinations.  We seek to provoke
exit relays to tamper with our connections, thus revealing their malicious intent.  By doing so, we
hope to discover and remove all ``spoiled onions'' which might be part of the Tor network.

We will also show that our scanner's \emph{modular design} enables quick prototyping of new scanning
modules.  Also, its \emph{event-driven architecture} makes it possible to scan the entire Tor
network within a matter of only seconds while at the same time sparing its resources.

\subsection{The Design of \textsf{exitmap}}
\label{sec:scanner}
\begin{figure}[t]
	\centering
	\includegraphics[width=0.4\textwidth]{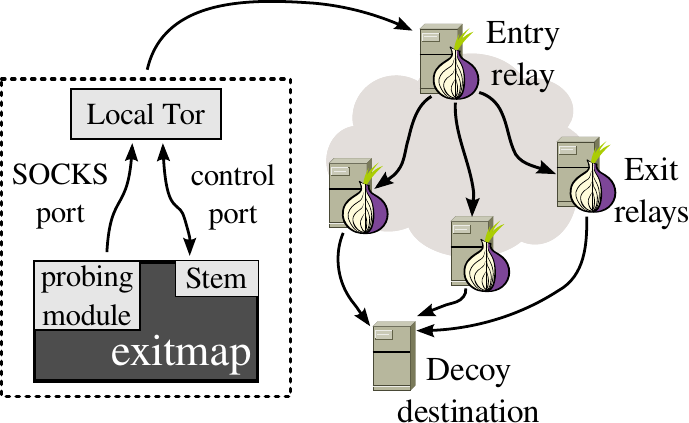}
	\caption{The design of \textsf{exitmap}.  Our scanner invokes a Tor process and uses the library
	\textsf{Stem} to control it.  Using \textsf{Stem}, circuits are created ``manually'' and
	attached to decoy connections which are initiated by our probing modules.}
	\label{fig:scanner}
\end{figure}

The schematic design of our scanner is illustrated in Figure~\ref{fig:scanner}.  Our tool is run on
a single machine and requires the Python library \textsf{Stem}~\cite{stem}.  \textsf{Stem}
implements the Tor control protocol~\cite{torcontrol} and we use it to initiate and close circuits,
attach streams to circuits as well as to parse the network consensus.  Upon starting
\textsf{exitmap}, it first invokes a local Tor process which proceeds by fetching the newest network
consensus in order to know which exit relays are currently online.

Next, our tool is fed with a set of exit relays.  This set can consist of a single relay, all exit
relays in a given country, or the set of all Tor exit relays.  Random permutation is then performed
on the set so that repeated scans do not probe exit relays in the same order.  This is useful
while developing and debugging new scanning modules as it equally distributes the load over all
selected exit relays.

Once \textsf{exitmap} knows which exit relays it has to probe, it initiates circuits which use the
respective exit relays as last hop.  All circuits are created asynchronously in the background.
Once a circuit to an exit relay is established, Tor informs \textsf{exitmap} about the circuit by
sending an asynchronous circuit event over the control connection.  Upon receiving the notification
about a successfully created circuit, \textsf{exitmap} invokes the desired probing module which then
proceeds by establishing a connection to a decoy destination (see \S~\ref{sec:modules}).  Tor
creates stream events for new connections to the SOCKS port which are also sent to \textsf{exitmap}.
At this point, we attach the stream of a probing module to the respective circuit.  Note that
stream-to-circuit attaching is typically done by Tor.  In order to have control over this action,
our scanner invokes Tor with the configuration option \texttt{\_\_LeaveStreamsUnattached} which
instructs Tor to leave streams unattached.

For performance reasons, Tor builds circuits preemptively, i.e., a number of circuits are kept ready
even if there is no data to be sent yet.  Since we want full control over all circuits, we prevent
Tor from creating circuits preemptively by using the configuration option
\texttt{\_\_DisablePredictedCircuits}.

Probing modules can either be standalone processes or Python modules.  Processes are invoked over
the \texttt{torsocks} wrapper \cite{torsocks} which hijacks system calls such as \texttt{socket()},
\texttt{connect()}, and \texttt{gethostbyname()} in order to redirect them to Tor's SOCKS port.  We
used standalone processes for our HTTPS and SSH modules.  In addition, probing modules can be
implemented in Python.  To redirect Python's networking API over Tor's SOCKS port, we extended the
SocksiPy module \cite{socksipy}.  We used Python for our sslstrip and DNS modules.

\subsection{Performance Hacks}
A naive approach to probing exit relays could cause non-trivial costs for the Tor network; mostly
computationally but also in terms of network throughput.  We implemented a number of tweaks in order
for our scanning to be as fast and cheap as possible.

\begin{figure}[t]
	\centering
	\includegraphics[width=0.45\textwidth]{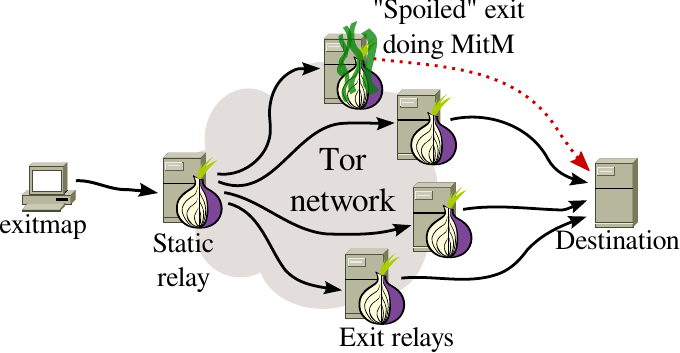}
	\caption{Instead of establishing a full three-hop circuit, our scanner is able to use a static
		middle relay; preferably operated by whoever is running our scanner.  By doing so, we
		concentrate the load on one machine while making our scanning activity slightly more
		obvious.}
	\label{fig:twohops}
\end{figure}

First, we expose a configuration option for avoiding the default of three-hop circuits.  Instead, we
only use \emph{two hops} as illustrated in Figure~\ref{fig:twohops}.  Tor's motivation for three
hops is anonymity but since our scanner has no need for strong anonymity, we only select a static
entry relay---ideally operated by \textsf{exitmap}'s user---which then directly forwards all traffic
to the respective exit relays.  We offer no option to use one-hop circuits as that would make it
possible for exit relays to isolate scanning connections:  A malicious exit relay could decide not
to tamper with a circuit if it originates from a non-Tor machine.  Since we use a static first hop
which is operated by us, we concentrate most of the scanning load on a single machine which is
well-suited to deal with the load.  Other entry and middle relays do not have to ``suffer'' from
scans.  However, note that over time malicious exit relays are able to correlate scans with relays,
thus determining which relays are used for scans.  To avoid this problem, \textsf{exitmap}'s first
hop could be changed periodically and we hope that by crowd-sourcing our scanner, isolating middle
relays is no longer a viable option for attackers.

Another computational performance tweak can be achieved on Tor's authentication layer.  At the
moment, there are two ways how a circuit handshake can be conducted; either by using the
\emph{traditional TAP} or the \emph{newer NTor} handshake.  TAP---short for Tor Authentication
Protocol~\cite{Goldberg2006}---is based on Diffie-Hellman key agreement in a multiplicative group.
NTor, on the other hand, uses the more efficient elliptic curve group
Curve25519~\cite{Bernstein2006}.  A non-trivial fraction of a relay's computational load can be
traced back to computationally expensive circuit handshakes.  By preferring NTor over TAP, we
slightly reduce the computational load on exit relays.  Since NTor supersedes TAP and is becoming
more and more popular as Tor clients upgrade, we believe that it is not viable for attackers to
``whitelist'' NTor connections.


\subsection{Scanning Modules}
\label{sec:modules}
After discussing the architecture of \textsf{exitmap}, we now present several probing modules we
developed in order to detect specific attacks.  When designing a module, it is important to consider
its \emph{indistinguishability} from genuine Tor clients.  As mentioned above, malicious relay
operators could closely inspect exit traffic (e.g., by examining the user agent string of browsers)
and only attack connections which appear to be genuine Tor users.

\subsubsection{HTTPS}
McCoy \emph{et al.}~\cite{McCoy2008} showed that HTTP is the most popular protocol in the Tor
network, clearly dominating other protocols such as instant messaging or e-mail\footnote{This is
particularly true based on \emph{connections} but not so much based on \emph{bytes
transferred}.}.  While HTTPS lags behind, it is still widely used and unsurprisingly, several exit
relays were documented to have tampered with HTTPS connections~\cite{badexits} in the past.

We implemented an HTTPS module which fetches a decoy destination's X.509 certificate and extracts
its fingerprint.  This fingerprint is then compared to the expected fingerprint which is hard-coded
in the module.  If there is a mismatch, an alert is triggered.  Originally, we began by fetching the
certificate using the command line utility \textsf{gnutls-cli}.  We later extended the module to
send a TLS client hello packet as it is sent by TorBrowser to make the scan less distinguishable
from what a real Tor user would send.

Note that an attacker might become suspicious after observing that a Tor user only fetched an X.509
certificate without actually browsing the web site.  However, at the point in time an attacker would
become suspicious, we already have what we need; namely the X.509 certificate.  Also, our module
could be extended to simulate simple browsing activity.

\subsubsection{sslstrip}
Instead of \emph{interfering} with TLS connections, an attacker can seek to \emph{prevent} TLS
connections.  This is the purpose of the tool \textsf{sslstrip}~\cite{sslstrip}.  The tool achieves
this goal by transparently rewriting HTML documents sent from the server to the client.  In
particular, it rewrites HTTPS links to HTTP links.  A secure login form such as
\texttt{https://login.example.com} is subsequently rewritten to HTTP which can cause a user's
browser to submit her credentials in the clear.  While the HTTP Strict Transport Security
policy~\cite{rfc6797} prevents sslstrip, it is still an effective attack against many large-scale
web sites with Yahoo! being one of them as of January 2014.  From an attacker's point of view, the
benefit of \textsf{sslstrip} is that it is a comparatively silent attack.  Browsers will not show
certificate warnings but vigilant users might notice the absence of browser-specific TLS
indicators such as a green address bar.

We implemented a probing module which can detect \textsf{sslstrip} attacks.  Our module fetches web
sites containing HTTPS links over unencrypted HTTP.  Afterwards, the module simply verifies whether
the fetched HTML document contains the expected HTTPS links or if they were ``downgraded'' to HTTP.
After experiments in a lab setting showed our module to work, we began \textsf{sslstrip} scans on
October 24, 2013.

\subsubsection{SSH}


The Tor network is also used to transport SSH traffic.  This can easily be done with the help of
tools such as \textsf{torsocks}~\cite{torsocks}.  Analogous to HTTPS-based attacks, malicious exit
relays could run MitM attacks against SSH.  In practice, this is not as easy as targeting HTTPS
given SSH's ``trust on first use'' model.  As long as the very first connection to an SSH server
with a given key was secure, the public key is then stored by the client and kept as reference for
subsequent connections.  That way, SSH is able to print a warning whenever the server's public key
is unexpected.  As a result, a MitM attack has to target a client's very first SSH connection where
the server's public key is not yet known.

\begin{figure}
\begin{lstlisting}
function probe( fingerprint, command ) {

    ssh_public_key = "11:22:33:44:55:66:77:88" +
                     "99:00:aa:bb:cc:dd:ee:ff";

    output = command.execute("ssh -v 1.2.3.4");

    if (ssh_public_key not in output) {
        print("Possible MitM attack by " + fingerprint);
    }
}
\end{lstlisting}
\caption{Pseudo code illustrating a scanning module which tests SSH.  It establishes an SSH
	connection to a given host and verifies if the fingerprint is as expected.  If the observed
	fingerprint differs, an alert is raised.}
\label{lst:ssh}
\end{figure}

Nevertheless, this practical problem might not stop attackers from attempting to interfere with SSH
connections.  Our SSH module, conceptually similar to the pseudo code shown in Figure~\ref{lst:ssh},
makes use of OpenSSH's \textsf{ssh} and \textsf{torsocks} to connect to a decoy server.  Again, the
server's key fingerprint is extracted and compared to the hard-coded fingerprint.  However, compared
to the HTTPS module, it is difficult to achieve indistinguishability over time.  After all, a
malicious relay operator could monitor an entire SSH session.  If it looks suspicious, e.g., it only
fetches the public key, or it lasts only one second, the attacker could decide to whitelist the
destination in the future.  Alternatively, we could establish SSH connections to random hosts on the
Internet.  This, however, is often considered undesired scanning activity and does not constitute
good Internet citizenship.  Instead, we again seek to solve this problem by publishing our source
code and encouraging people to crowdsource \textsf{exitmap} scanning.  Every \textsf{exitmap} user
is encouraged to use her own SSH server as decoy destination.  That way, we can achieve destination
diversity without bothering arbitrary SSH servers on the Internet.

\subsubsection{DNS}
While the Tor protocol only transports TCP streams, clients can ask exit relays to do DNS resolution
by wrapping domain names in a \texttt{RELAY\_BEGIN} cell~\cite{torspec}.  This cell is then sent to
the exit relay, once a circuit was established.  In the past, some exit relays were found to
inadvertently censor DNS queries, e.g., by using an OpenDNS configuration which blocks certain
domain categories such as ``Pornography'' or ``Proxy/Anonymiser''~\cite{badexits}.  Recall that
while such behaviour is not intentionally malicious, it is certainly enough to get the BadExit flag
assigned.

Our probing module maintains a whitelist of domains together with their corresponding IP addresses
and raises an alert if the DNS A record of a domain name is unexpected.  This approach works well
for sites with a known set of IP addresses but large sites frequently employ a diverse---and
sometimes geographically load-balanced---set of IP addresses which is difficult to enumerate.  Our
module probes several domains in the categories finance, social networking, political activism, and
pornography.

%

\subsection{Ethical Considerations}
Due to \textsf{exitmap}'s modular architecture, it can be used for various unintended---and even
unethical---purposes.  For example, modules for web site scraping or online voting manipulation come
to mind.  All sites which naively bind identities to IP addresses might be an attractive target.
While we do not endorse such actions, we point out that these activities are hard to stop and will
continue to happen and already happen regardless; with or without scanner.  If somebody decides to
abuse our scanner for such actions, it will at least spare the Tor network's resources more than a
naive design.  As a result, we believe that by publishing our code, the benefit to the public
outweighs the damage caused by unethical use.

\section{Experimental Results}
\label{sec:results}
On September 19th, we ran our first full scan over all $\sim$950 exit relays which were part of the
Tor network at the time. From then on, we scanned all exit relays several times a week.  Originally,
we began our scans while only armed with our HTTPS module but as time passed, we added additional
modules which allowed us to scan for additional attacks.  In this section, we will discuss the
results we obtained by monitoring the Tor network over a period of several months.

\subsection{Scanning Performance}
\begin{figure}[t]
	\centering
	\includegraphics[width=0.35\textwidth]{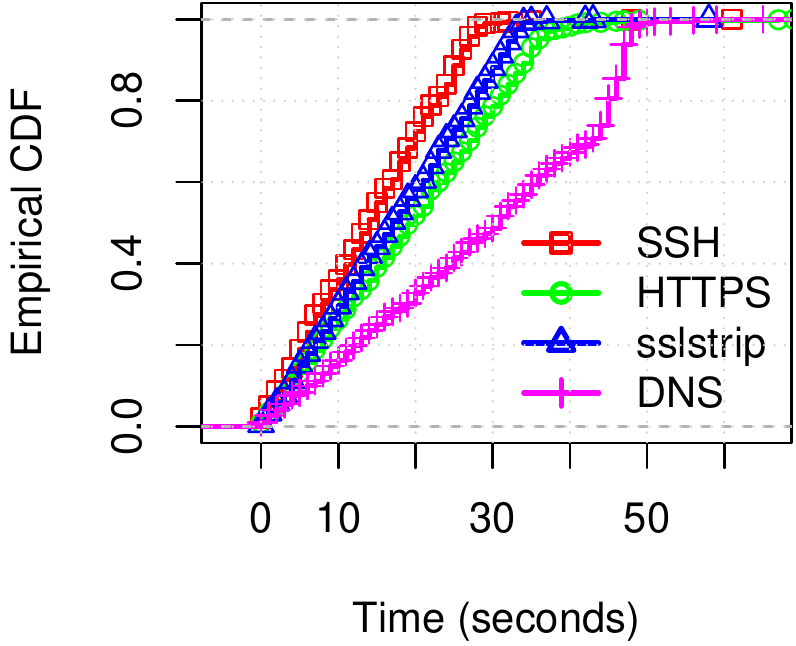}
	\caption{The performance of our probing modules.  The DNS module is slower because it resolves
		several domain names at once.  All other modules can scan at least 98\% of all Tor exit
		relays under 40 seconds.}
	\label{fig:performance}
\end{figure}

The performance of our probing modules is illustrated in Figure~\ref{fig:performance}.  The ECDF's
$x$-axis shows the time it takes for a module to finish successfully.  The $y$-axis shows the
cumulative fraction of all exit relays.  The diagram shows that all modules are able to scan at
least 98\% of all Tor exit relays  under 50 seconds.

Our data further shows that for all modules, 84\%--88\% of circuit creations succeeded.  The
remaining circuits either timed out or were torn down by the respective exit relay using a
\texttt{DESTROY} cell.

\subsection{Malicious Relays}
Table~\ref{tab:scans} contains the 25 malicious and misconfigured exit relays we found.  We
discovered the first two relays ``manually'' before we had developed \textsf{exitmap}.  All the data
illustrated in the table was gathered on the day we found the respective attack.  The columns are,
from left to right:
\begin{description}
	\item[\textbf{Fingerprint}] The first 4 bytes of the relay's unique 20-byte SHA-1 fingerprint.
	\item[\textbf{IP addresses}] All IPv4 addresses or netblocks, the relay was found to have used
		over its life time.
	\item[\textbf{Country}] The country in which the relay resided.  The country was
		determined with the help of MaxMind's GeoIP lite database.
	\item[\textbf{Bandwidth}] The advertised bandwidth, the relay was willing to contribute to the
		network.
	\item[\textbf{Attack}] The attack, the relay was running or its configuration problem.
	\item[\textbf{Sampling rate}] The sampling rate of the attack, i.e., how many connections were
		affected.
	\item[\textbf{First active}] The day, the relay was set up.
	\item[\textbf{Discovery}] The day, we discovered the relay.
\end{description}

\begin{table*}
\scriptsize
\caption{All 25 malicious and misconfigured exit relays we discovered over a period of 4 months.
The data was collected right after a relay was discovered.  We have reason to believe that all
relays whose fingerprint ends with a \dag\ were run by the same attacker.}
\label{tab:scans}
\centering
\begin{tabular}{lccccccc}
	\textbf{Fingerprint} & \textbf{IP addresses} & \textbf{Country} & \textbf{Bandwidth} &
	\textbf{Attack} & \textbf{Sampling rate} & \textbf{First active} & \textbf{Discovery} \\
\hline

\texttt{F8FD29D0}\dag & 176.99.12.246 & Russia & 7.16 MB/s & HTTPS MitM & \emph{unknown} & 2013-06-24 &
2013-07-13 \\
\hline

\texttt{8F9121BF}\dag & 64.22.111.168/29 & U.S. & 7.16 MB/s & HTTPS MitM & \emph{unknown} &
2013-06-11 & 2013-07-13 \\
\hline
\hline

\texttt{93213A1F}\dag & 176.99.9.114 & Russia & 290 KB/s & HTTPS MitM & 50\% & 2013-07-23 & 2013-09-19 \\
\hline

\texttt{05AD06E2}\dag & 92.63.102.68 & Russia & 5.55 MB/s & HTTPS MitM & 33\% & 2013-08-01 & 2013-09-19 \\
\hline

\texttt{45C55E46}\dag & 46.254.19.140 & Russia & 1.54 MB/s & SSH \& HTTPS MitM & 12\% & 2013-08-09 &
2013-09-23 \\
\hline

\texttt{CA1BA219}\dag & 176.99.9.111 & Russia & 334 KB/s & HTTPS MitM & 37.5\% & 2013-09-26 & 2013-10-01 \\
\hline

\texttt{1D70CDED}\dag & 46.38.50.54 & Russia & 929 KB/s & HTTPS MitM & 50\% & 2013-09-27 & 2013-10-14 \\
\hline

\texttt{EE215500}\dag & 31.41.45.235 & Russia & 2.96 MB/s & HTTPS MitM & 50\% & 2013-09-26 & 2013-10-15 \\
\hline

\texttt{12459837}\dag & 195.2.252.117 & Russia & 3.45 MB/s & HTTPS MitM & 26.9\% & 2013-09-26 & 2013-10-16
\\
\hline

\texttt{B5906553}\dag & 83.172.8.4 & Russia & 850.9 KB/s & HTTPS MitM & 68\% & 2013-08-12 & 2013-10-16 \\
\hline

\texttt{EFF1D805}\dag & 188.120.228.103 & Russia & 287.6 KB/s & HTTPS MitM & 61.2\% & 2013-10-23 &
2013-10-23 \\
\hline

\texttt{229C3722} & 121.54.175.51 & Hong Kong & 106.4 KB/s & sslstrip & \emph{unsampled} & 2013-06-05 &
2013-10-31 \\
\hline

\texttt{4E8401D7}\dag & 176.99.11.182 & Russia & 1.54 MB/s & HTTPS MitM & 79.6\% & 2013-11-08 & 2013-11-09
\\
\hline

\texttt{27FB6BB0}\dag & 195.2.253.159 & Russia & 721 KB/s & HTTPS MitM & 43.8\% & 2013-11-08 & 2013-11-09 \\
\hline

\texttt{0ABB31BD}\dag & 195.88.208.137 & Russia & 2.3 MB/s & SSH \& HTTPS MitM & 85.7\% & 2013-10-31 &
2013-11-21 \\
\hline

\texttt{CADA00B9}\dag & 5.63.154.230 & Russia & 187.62 KB/s & HTTPS MitM & \emph{unsampled} & 2013-11-26 &
2013-11-26 \\
\hline

\texttt{C1C0EDAD}\dag & 93.170.130.194 & Russia & 838.54 KB/s & HTTPS MitM & \emph{unsampled} &
2013-11-26 & 2013-11-27 \\ \hline

\texttt{5A2A51D4} & 111.240.0.0/12 & Taiwan & 192.54 KB/s & HTML
Injection & \emph{unsampled} & 2013-11-23 & 2013-11-27 \\
\hline

\texttt{EBF7172E}\dag & 37.143.11.220 & Russia & 4.34 MB/s & SSH MitM & \emph{unsampled} &
2013-11-15 & 2013-11-27 \\
\hline

\texttt{68E682DF}\dag & 46.17.46.108 & Russia & 60.21 KB/s & SSH \& HTTPS MitM & \emph{unsampled} &
2013-12-02 & 2013-12-02 \\
\hline

\texttt{533FDE2F}\dag & 62.109.22.20 & Russia & 896.42 KB/s & SSH \& HTTPS MitM & 42.1\% &
2013-12-06 & 2013-12-08 \\
\hline

\texttt{E455A115} & 89.128.56.73 & Spain & 54.27 KB/s & sslstrip & \emph{unsampled} & 2013-12-17
& 2013-12-18 \\
\hline

\texttt{02013F48} & 117.18.118.136 & Hong Kong & 538.45 KB/s & DNS censorship & \emph{unsampled} &
2013-12-22 & 2014-01-01 \\
\hline

\texttt{2F5B07B2} & 178.211.39 & Turkey & 204.8 KB/s & DNS censorship & \emph{unsampled} &
2013-12-28 & 2014-01-06 \\
\hline

\texttt{4E2692FE} & 24.84.118.132 & Canada & 52.22 KB/s & OpenDNS & \emph{unsampled} & 2013-12-21 &
2014-01-06 \\
\hline

\end{tabular}
\end{table*}

Apart from all the conspicuous HTTPS MitM attacks which we will discuss later, we exposed two relays
running \textsf{sslstrip} for a short time.  The relay \texttt{5A2A51D4} injected custom HTML code
into HTTP traffic (see Appendix~\ref{app:injected_html}).  While the injected code seemed harmless
during our tests, we cannot rule out malicious intent.  Two more relays---\texttt{02013F48} and
\texttt{2F5B07B2}---were subject to their country's DNS censorship.  The Turkish relay blocked many
pornography web sites and redirected the user to a government-run web server which explained the
reason for the redirection.  The second relay seemed to have fallen prey to the Great Firewall of
China's DNS poisoning; perhaps, the relay made use of a DNS resolver in China.  Several domains such
as torproject.org, facebook.com and youtube.com returned invalid IP addresses which were also found
in previous work \cite{Lowe2007}.  Finally, \texttt{4E2692FE} was misconfigured because it used an
OpenDNS policy which would censor web sites in the category ``pornography''.

All the remaining relays engaged in HTTPS and/or SSH MitM attacks.  Upon establishing a connection
to the decoy destination, these relays exchanged the destination's certificate with their own,
self-signed version.  Since these certificates were not issued by a trusted authority contained in
TorBrowser's certificate store, a user falling prey to such a MitM attack would be redirected to the
\url{about:certerror} warning page.

Interestingly, we have reason to believe that all relays whose fingerprint ends with a \dag\ were
run by the same person or group of people.  This becomes evident when analysing the self-signed
certificates which were injected for the MitM attacks.  In every case, the certificate chain
consisted of only two nodes which both belonged to a ``Main Authority'' and the root
certificate---partially shown in Figure~\ref{lst:certificate}---of all chains was \emph{identical}.
This means that these attacks can be traced back to a common origin even though it is not clear
where or what this origin is as we will discuss later.

\begin{figure}[t]
\begin{lstlisting}
C=US
ST=Nevada
L=Newbury
O=Main Authority
OU=Certificate Management
CN=main.authority.com
EMAIL=cert@authority.com
\end{lstlisting}
\caption{X.509 information which is part of the malicious certificates used for the MitM attacks.
	The full certificate is shown in Appendix~\ref{app:x509}.}
\label{lst:certificate}
\end{figure}

Apart from the identical root certificate, these relays had other properties in common.  First, with
the exception of \texttt{8F9121BF} which was located in the U.S., they were \emph{all located in
Russia}.  Upon investigating their IP addresses, we discovered that most of the Russian relays were
run in the network of a virtual private system (VPS) provider.  Several IP addresses were also
located in the same netblock, namely 176.99.12.246, 176.99.9.114, 176.99.9.111, and 176.99.11.182.
All these IP addresses are part of the netblock GlobaTel-net which spans 176.99.0.0/20.
Furthermore, the malicious exit relays all used Tor version 0.2.2.37\footnote{For comparison, as of
January 2014, the current stable version is 0.2.4.20.  Version 0.2.2.37 was declared stable on June
6th, 2012.}.  Given its age, this is a rather uncommon version number amongst relays.  In fact, we
found only two benign exit relays---in Switzerland and the U.S.---which are running the same
version.  We suspect that the attackers might have a precompiled version of Tor which they simply
copy to newly purchased systems to spawn new exit relays.  Unfortunately, we have no data which
would allow us to verify when this series of attacks began.  However, the full root certificate
shown in Appendix~\ref{app:x509} indicates that it was created on February 12, 2013.

\subsection{Connection Sampling}
\label{sec:sampling}
Whenever our hunt for malicious relays yielded another result, we strived to confirm the attack by
rerunning the scan on the newly discovered relay.  However, in the case of the Russian relays, this
did not always result in the expected HTTPS MitM attack.  Instead, we found that only every $n$th
connection seemed to have been attacked.  We estimated the exact \emph{sampling rate} by
establishing 50 HTTPS connections over every relay.  We used randomly determined sleep
periods in between the scans in order to disguise our activity.  The estimated sampling rate for
every relay is shown in Table~\ref{tab:scans} in the column ``Sampling rate''.  For all Russian
relays, it varies between 12\% and 68\%.  We do not have an explanation for the attacker's
motivation to sample connections.  One theory is that sampling makes it less likely for a malicious
exit relay to be discovered; but at the cost of collecting fewer MitM victims.

Interestingly, the sampling technique was implemented \emph{ineffectively}.  This is due to the way
how Firefox (and as a result TorBrowser) reacts to self-signed certificates.  When facing a
self-signed X.509 certificate, Firefox displays its \url{about:certerror} page which warns the user
about the security risk.  If a user then decides to proceed, the certificate is \emph{fetched
again}.  We observed that the malicious exit relays treat the certificate re-fetching as a separate
connection whose success depends on the relay's sampling rate.  As a result, a sampling rate of $n$
means that a MitM attack will only be successfully with a probability of $n^2$.

\subsection{Who is the Attacker?}
An important question is where on the path from the exit relay to the destination the attacker is
located.  At first glance, one might blame the exit relay operator.  However, it is also possible
that the actual attack happens \emph{after} the exit relay, e.g., by the relay's ISP, the network
backbone, or the destination ISP.  In fact, such an incident was documented in 2006 for a relay
located in China \cite{upstream}.

With respect to our data, we cannot entirely rule out that the HTTPS MitM attacks were actually run
by an upstream provider of the Russian exit relays.  However, we consider it unlikely for the
following reasons: \emph{1)} the relays were located in diverse IP address blocks and there were
numerous other relays in Russia which did not exhibit this behaviour, \emph{2)} one of the relays
was even located in the U.S., \emph{3)} there are no other reported cases on the Internet involving
a certification authority called ``Main Authority'', and \emph{4)} the relays frequently disappeared
after they were assigned the BadExit flag.

The identity of the attacker is difficult to ascertain.  The relays did not publish any contact
information, nicknames, or revealed other hints which could enable educated guesses regarding the
attacker's origin.

\subsection{Destination Targeting}
While Tor's nature as an anonymity tool renders targeting individuals difficult\footnote{We assume,
of course, that users do not somehow reveal their real identity when using Tor, e.g., by posting on
Internet forums under their real name.}, an attacker can target classes of users based on their
communication \emph{destination}.  For example, an attacker could decide to only tamper with
connections going to the fictional www.insecure-bank.com.  Interestingly, we found evidence for
exactly that behaviour; at some point the Russian relays began to target at least
\url{facebook.com}.  We tested the https version of the Alexa top 10 web sites~\cite{alexa} but were
unable to trigger MitM attacks despite numerous connection attempts.  Popular Russian web sites such
as the mail provider \url{mail.ru} and the social network \url{vk.com} also remained unaffected.
Note that it is certainly possible that the relays targeted additional web sites we did not test
for.  Answering this question comprehensively would mean probing for thousands of different web
sites.

We have no explanation for the targeting of destinations.  It might be another attempt to delay the
discovery by vigilant users.  However, according to previous research~\cite{Huber2010}, social
networking appears to be as popular over Tor as it is on the clear Internet.  As a result, limiting
the attack to facebook.com might not significantly delay discovery.



\section{Thwarting HTTPS MitM Attacks}
\label{sec:thwarting}
The discovery of destination targeting made us reconsider defence mechanisms. Unfortunately, we
cannot rule out that there are additional, yet undiscovered exit relays which target low-profile web
sites.  If we wanted to achieve high coverage, we would have to connect to millions of web sites;
and given the connection sampling discussed in Section~\ref{sec:sampling}, this even has to be done
repeatedly!  After all, an attacker is able to \emph{arbitrarily reduce the scope} of the attack but
we are \emph{unable to arbitrarily scale} our scanner.  This observation motivated another defence
mechanism which is discussed in this section.

\subsection{Threat Model}
We consider an adversary who is controlling the upstream Internet connection of a small fraction of
exit relays\footnote{By ``fraction'', we mean a relay's bandwidth as it determines how likely a
client is to select the relay as part of its circuit.}. The adversary's goal is to run HTTPS-based
MitM attacks against Tor users.  We further expect the adversary to make an effort to stay under the
radar in order to delay discovery.  The actual MitM attack is conducted by injecting self-signed
certificates in the hope that users are not scared off by the certificate warning page.

Our threat model does not cover adversaries who control certificate authorities which would enable
them to issue valid certificates to avoid TorBrowser's warning page.  This includes several
countries as well as organisations which are part of TorBrowser's root certificate store.
Furthermore, we cannot defend against adversaries who control a significant fraction of Tor exit
bandwidth.

\subsection{Multi Circuit Certificate Verification}
As long as an attacker is unable to tamper with all connections to a given destination\footnote{This
would be the case if an attacker controls the destination.}, MitM attacks can be detected by
fetching a public key over \emph{differing paths in the network}.  This approach was picked up by
several projects including Perspectives~\cite{Wendlandt2008}, Convergence~\cite{convergence} and
Crossbear~\cite{crossbear}. In this section, we discuss a patch for TorBrowser which achieves the
same goal but is adapted to the Tor network.

Apart from NoScript and HTTPS-Everywhere, TorBrowser contains another important extension:
\emph{Torbutton}.  This extension provides the actual interface between TorBrowser and the local Tor
process.  It directs TorBrowser's traffic to Tor's SOCKS port and exposes a number of features such
as the possibility to create a new identity.

Torbutton already contains rudimentary code to talk to Tor over the local control port.  The
control port---typically bound to 127.0.0.1:9151---provides local applications with an interface to
control Tor.  For example, Torbutton's ``New Identity'' feature works by sending the \texttt{NEWNYM}
signal which instructs Tor to switch to clean circuits so that new application requests do not share
circuits with old requests.  Torbutton already implements a useful code base for us which made us
decide to implement our extension as a set of patches for Torbutton rather than build an independent
extension.

\subsection{Extension Design}
Our patch set hooks into the browser event \texttt{DOMContentLoaded} which is triggered whenever a
document (but not necessarily stylesheets and images) is loaded and parsed by the browser.  We
then check if the URI of the page contains ``\url{about:certerror}'' because whenever TorBrowser
encounters a self-signed certificate, it displays this page.  However, it is not clear whether the
certificate is genuinely self-signed or part of an attack.

\begin{figure}[t]
	\centering
	\includegraphics[width=0.47\textwidth]{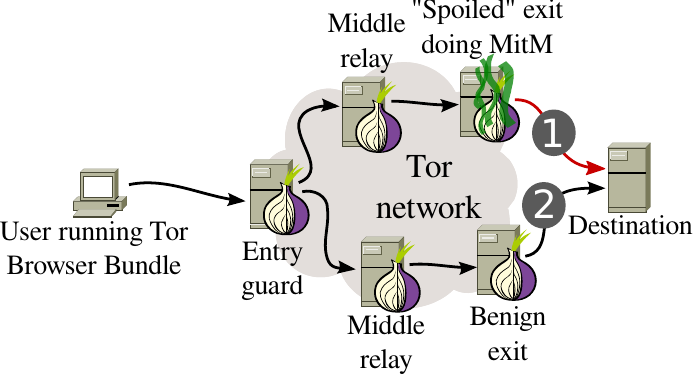}
	\caption{A user stumbles across a self-signed certificate \ding{202} which could be an
	indication for a HTTPS MitM attack ran by a malicious exit relay.  To verify if the certificate
	is genuine, the client re-fetches it over an independent exit relay \ding{203} and checks if
	the certificate matches or
	not.}
	\label{fig:verification}
\end{figure}

In order to be able to distinguish between these two cases, our patch now attempts to re-fetch the
certificate over at least one additional and distinct Tor circuit as illustrated in
Figure~\ref{fig:verification}.  We create a fresh circuit by sending \texttt{SIGNAL NEWNYM} to Tor's
control port.  Afterwards, we re-fetch the certificate by issuing an \texttt{XMLHttpRequest}.  If
the SHA-1 fingerprints of both certificates match, the certificate is probably\footnote{Note that
powerful adversaries might be able to control multiple exit relays, network backbones, or even the
destination.} genuine.  Otherwise, the user might have fallen prey to a MitM attack.
False positives are possible, though: large sites could have different certificates for different
geographical regions.  Note that we are not very likely to witness many such false positives as our
code only gets active upon observing self-signed certificates or certificates which somehow trigger
the \url{about:certerror} warning page.

\begin{figure}[t]
	\centering
	\includegraphics[width=0.47\textwidth]{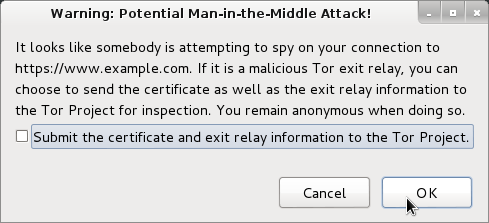}
	\caption{The popup window in TorBrowser which informs the user about the potential HTTPS MitM
	attack.  The user can agree to submitting the gathered information to the Tor Project for
	further inspection.}
	\label{fig:popup}
\end{figure}

Our extension also informs the user about a potential MitM attack: In case of differing
certificates, we open a browser dialogue which informs the user about the situation.  A screenshot
of our design prototype is shown in Figure~\ref{fig:popup}.  We point out that this is likely an
attack and we ask the user for permission to send the data to the Tor Project for further
inspection.  The submitted data contains the \emph{exit relays} used for certificate fetching as
well as the \emph{observed certificates}.  We transmit no other data which could be used to identify
users; as a result, certificate submission is anonymous.  While it would be technically possible to
transmit the data silently, we believe that users would not appreciate it and consider it as
``phoning home''.  As a result, we seek to obtain informed consent.

\subsection{Limitations}
In our threat model, we mentioned that our design does not protect against adversaries with the
ability to issue valid certificates.  While our extension could easily be extended to conduct
certificate comparison for all observed certificates, it would flood the Tor network with
certificate re-fetches.  To make matters worse, the overwhelming majority of these re-fetches would
not even expose attacks.  There exist other techniques to foil CA-capable adversaries such as
certificate pinning~\cite{pinning}.

By default, our patch re-fetches a self-signed X.509 certificate only once.  An attacker who is
controlling a significant fraction of exit relays might be able to conduct a MitM attack for the
first as well as for the second fetch.  Nevertheless, we would eventually catch the adversary; it
would simply be a matter of time until a user selects two independent exit relays.

\section{Conclusions}
\label{sec:conclusions}
In this paper, we revisited the trustworthiness of Tor exit relays.  After developing a scanner, we
closely monitored all $\sim$1,000 exit relays over a period of four months.  We discovered 25 relays
which were either outright malicious or simply misconfigured.  Interestingly, the majority of the
attacks were coordinated instead of being isolated actions of independent individuals.  Our results
further suggest that the attackers make an active effort to remain under the radar and delay
detection.

To make the Tor network safer, we first developed \textsf{exitmap}; an easily extensible scanner
which is able to probe exit relays for a variety of MitM attacks.  Furthermore, we developed a set
of patches for the Tor Browser Bundle which is capable of fetching self-signed X.509 certificates
over different network paths to evaluate their trustworthiness.  We believe that by being armed with
these two tools, the security of the Tor network can be greatly increased.  Finally, all our source
code is freely available: \url{http://www.cs.kau.se/philwint/spoiled_onions}.

\section*{Acknowledgements}
We want to thank Internetfonden whose research grant made this work possible.  Furthermore, we want
to thank Aaron Gibson, Georg Koppen, Harald Lampesberger, and Linus Nordberg for helpful feedback
and suggestions.

\printbibliography

\appendix

\section{Malicious X.509 Root Certificate}
\label{app:x509}
Below, the root certificate which was shared by all Russian and the single U.S. exit relay is shown.
While the domain \url{authority.com} does exist, it seems unrelated to the CA ``Main Authority'',
the issuer.


\begin{lstlisting}[basicstyle=\scriptsize\ttfamily]
Certificate:
 Data:
  Version: 3 (0x2)
  Serial Number: 16517615612733694071 (0xe53a5be2bd702077)
 Signature Algorithm: sha1WithRSAEncryption
  Issuer: C=US, ST=Nevada, L=Newbury, O=Main Authority,
    OU=Certificate Management,
    CN=main.authority.com/emailAddress=cert@authority.com
  Validity
   Not Before: Feb 12 08:13:07 2013 GMT
   Not After : Feb 10 08:13:07 2023 GMT
  Subject: C=US, ST=Nevada, L=Newbury, O=Main Authority,
     OU=Certificate Management,
     CN=main.authority.com/emailAddress=cert@authority.com
  Subject Public Key Info:
   Public Key Algorithm: rsaEncryption
    Public-Key: (1024 bit)
    Modulus:
     00:da:5d:5f:06:06:dc:8e:f1:8c:70:b1:58:12:0a:
     41:0e:b9:23:cc:0e:6f:bc:22:5a:05:12:09:cf:ac:
     85:9d:95:2c:3a:93:5d:c9:04:c9:4e:72:15:6a:10:
     f1:b6:cd:e4:8e:ad:5a:7f:1e:d2:b5:a7:13:e9:87:
     d8:aa:a0:24:15:24:84:37:d1:69:8e:31:8f:5c:2e:
     92:e3:f4:9c:c3:bc:18:7d:cf:b7:ba:b2:5b:32:61:
     64:05:cd:1f:c3:b5:28:e1:f5:a5:1c:35:db:0f:e8:
     c3:1d:e3:e3:33:9c:95:61:6d:b7:a6:ad:de:2b:0d:
     d2:88:07:5f:63:0d:9c:1e:cf
    Exponent: 65537 (0x10001)
  X509v3 extensions:
   X509v3 Subject Key Identifier: 
    07:42:E0:52:A7:DC:A5:C5:0F:C5:
    AF:03:56:CD:EB:42:8D:96:00:D6
   X509v3 Authority Key Identifier: 
    keyid:07:42:E0:52:A7:DC:A5:C5:0F:C5:
          AF:03:56:CD:EB:42:8D:96:00:D6
    DirName:/C=US/ST=Nevada/L=Newbury/O=Main Authority
      /OU=Certificate Management
      /CN=main.authority.com/emailAddress=cert@authority.com
    serial:E5:3A:5B:E2:BD:70:20:77

   X509v3 Basic Constraints: 
    CA:TRUE
 Signature Algorithm: sha1WithRSAEncryption
   23:55:73:1b:5c:77:e4:4b:14:d7:71:b4:09:11:4c:ed:2d:08:
   ae:7e:37:21:2e:a7:a0:49:6f:d1:9f:c8:21:77:76:55:71:f9:
   8c:7b:2c:e8:a9:ea:7f:2f:98:f7:45:44:52:b5:46:a4:09:4b:
   ce:88:90:bd:28:ed:05:8c:b6:14:79:a0:f3:d3:1f:30:d6:59:
   5c:dd:e6:e6:cd:3a:a4:69:8f:2d:0c:49:e7:df:01:52:b3:34:
   38:97:c5:9a:c3:fa:f3:61:b8:89:0f:d2:d9:a5:48:e6:7b:67:
   48:4a:72:3f:da:28:3e:65:bf:7a:c2:96:27:dd:c0:1a:ea:51:
   f5:09
\end{lstlisting}

\section{Injected HTML Code}
\label{app:injected_html}
The following HTML code was injected by the relay \texttt{5A2A51D4} (see Table~\ref{tab:scans}).  It
was appended right in front of the closing HTML tag.

\begin{lstlisting}
<br>
<img src="http://111.251.157.184/pics.cgi"
 width="1" height="1">
\end{lstlisting}

When requesting the image link inside the HTML code, the server responds with another HTML document.
The full HTTP response is shown below.

\begin{lstlisting}
HTTP/1.1 200 OK
Date: Tue, 14 Jan 2014 17:12:08 GMT
Server: Apache/2.2.22 (Ubuntu)
Vary: Accept-Encoding
Transfer-Encoding: chunked
Content-Type: text/html


<HTML>
<HEAD>
<TITLE>No Title</TITLE>
</HEAD>
<BODY>

</BODY>
</HTML>
\end{lstlisting}

\end{document}